\documentclass[superscriptaddress,pre,reprint,showpacs]{revtex4-1}

\usepackage{amsmath,empheq, array, bigints}
\usepackage{bm}
\usepackage{amsfonts}
\usepackage{amsthm}
\usepackage{amssymb}
\usepackage{graphicx}
\usepackage{hyperref}
\usepackage{float}
\usepackage[format=plain,justification=justified]{caption}
\usepackage[format=plain,justification=justified]{subcaption}
 
\usepackage{soul}

\newcommand{\ncd}{\newcommand}
\ncd{\mrm}    {\mathrm}
\ncd{\beq} {\begin{equation}}
  \ncd{\eeq} {\end{equation}}

\def\E{\mathbb{E}}

\def\d{{\rm d}}

\def\r{{\bm{r}}}

\newcommand\norm[1]{\left\lVert#1\right\rVert}

\def\avg#1{\left< #1 \right>}

\graphicspath{{./figures/}}

\begin{document}

       \title{Monte Carlo sampling in diffusive dynamical systems}

	\date{\today}

	\author{Diego Tapias}
	\email{diego.tapias@nucleares.unam.mx }
        \affiliation{Departamento de F\'isica, Facultad de Ciencias, Universidad Nacional Aut\'onoma de M\'exico, Ciudad Universitaria, Ciudad de M\'exico 04510, Mexico}
        \affiliation{School of Mathematics and Statistics, University of Sydney, 2006, NSW, Sydney, Australia}

	\author{David P. Sanders }
        \email{dpsanders@ciencias.unam.mx}
        \affiliation{Departamento de F\'isica, Facultad de Ciencias, Universidad Nacional Aut\'onoma de M\'exico, Ciudad Universitaria, Ciudad de M\'exico 04510, Mexico}

	\author{Eduardo G. Altmann}
        \email{eduardo.altmann@sydney.edu.au}
        \affiliation{School of Mathematics and Statistics, University of Sydney, 2006, NSW, Sydney, Australia}

	\begin{abstract}

          We introduce a Monte Carlo algorithm to efficiently compute transport properties of chaotic dynamical systems.
          Our method exploits the importance sampling technique that favors trajectories in the tail of the distribution of displacements, where deviations from a diffusive process are most prominent. We search for initial conditions using a proposal that correlates states in the Markov chain constructed via a Metropolis-Hastings algorithm. We show that our method outperforms the direct sampling method and also Metropolis-Hastings methods with alternative proposals. We test our general method through numerical simulations in 1D (box-map) and 2D (Lorentz gas) systems. \\
\end{abstract}


\maketitle

\noindent{\bf The statistical description of physical systems, at first justified based on their high dimensionality, is now well-known to be applicable and necessary already in simple, low-dimensional, chaotic systems. Even though analytical results are often hard to obtain, numerical simulations in broad classes of sytems confirm that the properties of ensembles of trajectories in deterministic systems are best described statistically. These numerical simulations become computationally expensive when the dimensionality increases or when the quantities of interest rely on atypical trajectories.  In this paper we develop numerical methods to efficiently simulate rare trajectories in chaotic systems showing diffusive behavior.}

\section{Introduction}

An example of a statistical description (at macroscopic scales) arising due to chaotic dynamics (at microscopic scales) is the onset of diffusion in different spatially extended {\it deterministic} systems~\cite{vollmer2002chaos,gaspard2005chaos, klages2007microscopic, klages2008anomalous, dorfman1999introduction}. Theoretically, these systems are in the diffusive regime in the limit of time going to infinity. The finite-time behavior, or the exact conditions of validity of the diffusive approximation, are typically unknown. In practice, one often relies on numerical simulations of an ensemble of trajectories to compute the distribution $\rho(r; t)$ of displacements $r$ at a given observation time $t$; see e.g.~Ref.~\cite{sanders2008deterministic} for the numerical analysis of the paradigmatic Lorentz gas. Such direct numerical simulations capture the behavior of typical trajectories, and are often adequate to estimate the diffusion coefficient; however, they struggle to characterize rare trajectories such as the ones in the tails of the distribution of displacements. 

Importance sampling is a variance reduction technique used for Monte Carlo calculations to compute rare configurations~\cite{rubino2009rare}. It relies on  the analysis of a weighted distribution of the random variable of interest, and is used either when the direct distribution is difficult to sample, or when the focus is on its tails. As a variance reduction technique it increases the precision of the estimates that can be obtained for a given computational effort~\cite{christian2004monte, rubino2009rare, bucklew2013introduction}. More recently, importance sampling methods have been increasingly applied to compute rare trajectories in  dynamical systems~\cite{tailleur2007probing,geiger2010identifying,leitao2017importance}, e.g. to study finite-time Lyapunov exponents~\cite{leitao2014efficiency, iba2014multicanonical}, open chaotic systems~\cite{lai2011transient},  climate systems~\cite{wouters2016rare}, chemical reactions~\cite{dellago2002transition, dellago2008transition}.

In this work we show how importance sampling methods can be applied to  compute transport properties of deterministic  dynamical systems. The displacement distribution $\rho(r; {t})$ is obtained through a Markov Chain Monte Carlo procedure, implemented via a Metropolis--Hastings algorithm. The key part of our method is an efficient proposal for moving within the set of initial conditions. By using information about the ensemble second moment, which is known to satisfy the standard diffusion rule, we deduce an algorithm that is applicable to study the tail of the displacement distribution. Our paper applies and extends the general framework introduced in Ref.~\cite{leitao2017importance}, correcting it to ensure the validity of the detailed-balance condition and extending its applicability to the computation of transport properties.

\section{Problem setting}

Consider a deterministic dynamical system~$\phi^t(x)$ that evolves initial conditions $x$ in the $d$-dimensional phase space $\Omega \subseteq \mathbb{R}^d$ for a time $t$, either discrete or continuous. The observable we are interested in is the (Euclidean norm of the) displacement  $  r_{t}(x) \equiv r(x, t) := \norm{\phi^{t}(x) - x}$ at an observation time $t$ and its distribution $\rho(r; {t})$, obtained from the set of initial conditions $\Gamma \subseteq \Omega$ and a naturally defined probability measure (e.g., uniform or the natural measure of $\phi$). The general problem we consider in this paper is how to efficiently estimate $\rho(r; {t})$ from the numerical simulation of a sample of $x\in\Gamma$. Transport properties are computed from $\rho(r; {t})$, e.g. the diffusion coefficient 
\beq
D := \lim_{t \rightarrow \infty} \frac{\avg{r^2}_{t}}{2 \bar{d} t} \, ,
\label{coefdif}
\eeq
where $\bar{d}$ is the spatial dimension and the mean of a spatial observable $f$ at time $t$ is defined as $\avg{f(r)}_{t} \equiv \int_0^\infty f(r) \rho(r; {t}) dr$.

The simplest, \emph{direct sampling}, approach consists of a uniform sampling of the space $\Gamma$ and the measurement of $r_{t}(x^{(i)})$ for each initial condition $x^{(i)}$. The estimator of the distribution $\rho(r; t)$ is then
\beq
\hat{\rho}(r; t) = \frac{1}{M} \sum_{i = 1}^M \delta(r - r_{t}(x^{(i)}) ) \, ,
\label{directrho}
\eeq
where $M$ is the number of initial conditions and $\delta$ is the Dirac delta distribution.
This method is appropriate to estimate the diffusion coefficient or typical trajectories of $\rho(r; t)$. However, the estimator~\eqref{directrho} requires a vast number of samples to be accurate in the tail or to calculate high moments of $\rho(r;t)$\cite{dettmann2014diffusion}. This is seen, for instance, in the problem of estimating the probability of finding trajectories above a given threshold $r > r^*$, e.g.~to estimate
\beq
I_{t} = \int_{r*}^\infty \rho(r; t) \, dr .
\label{integral}
\eeq
In this paper we develop a method suited to  problems in which such rare trajectories play an important role.

\subsection{Importance sampling}

Importance sampling methods are useful to reduce the computational cost of rare trajectories, e.g., reducing the scaling between estimators of Eq.~(\ref{integral}) and the number of samples $M$. This is achieved by sampling states with a weighted distribution, different from $\rho(r;t)$. Here we focus on the canonical (or escort) distribution~\cite{beck1995thermodynamics}
\beq
\rho_{\beta}(r; t) = \frac{\rho(r; t ) \, {\rm e}^{-\beta r}}{Z_\beta} \, ,
\label{importancedistribution}
\eeq
where $Z_\beta$ denotes a normalization constant (partition function)
and $\beta$ (inverse temperature) is a parameter introduced to favor orbits in the tail of the distribution. 

\begin{figure*}
\begin{subfigure}{0.5\textwidth}
  \centering
  \includegraphics[width=\linewidth]{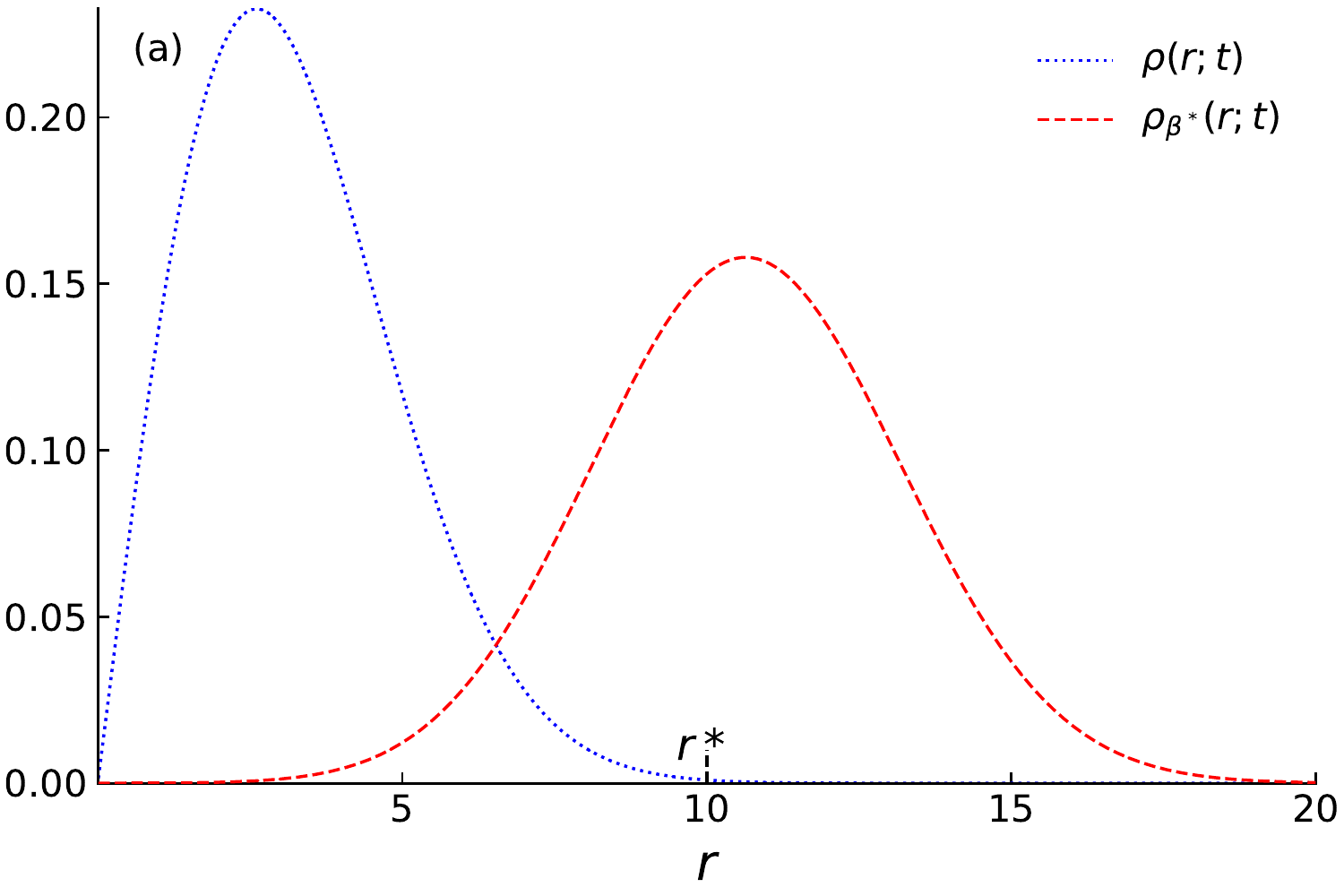}
  \label{rhoyrhobeta}
\end{subfigure}%
\begin{subfigure}{.5\textwidth}
  \includegraphics[width=\linewidth]{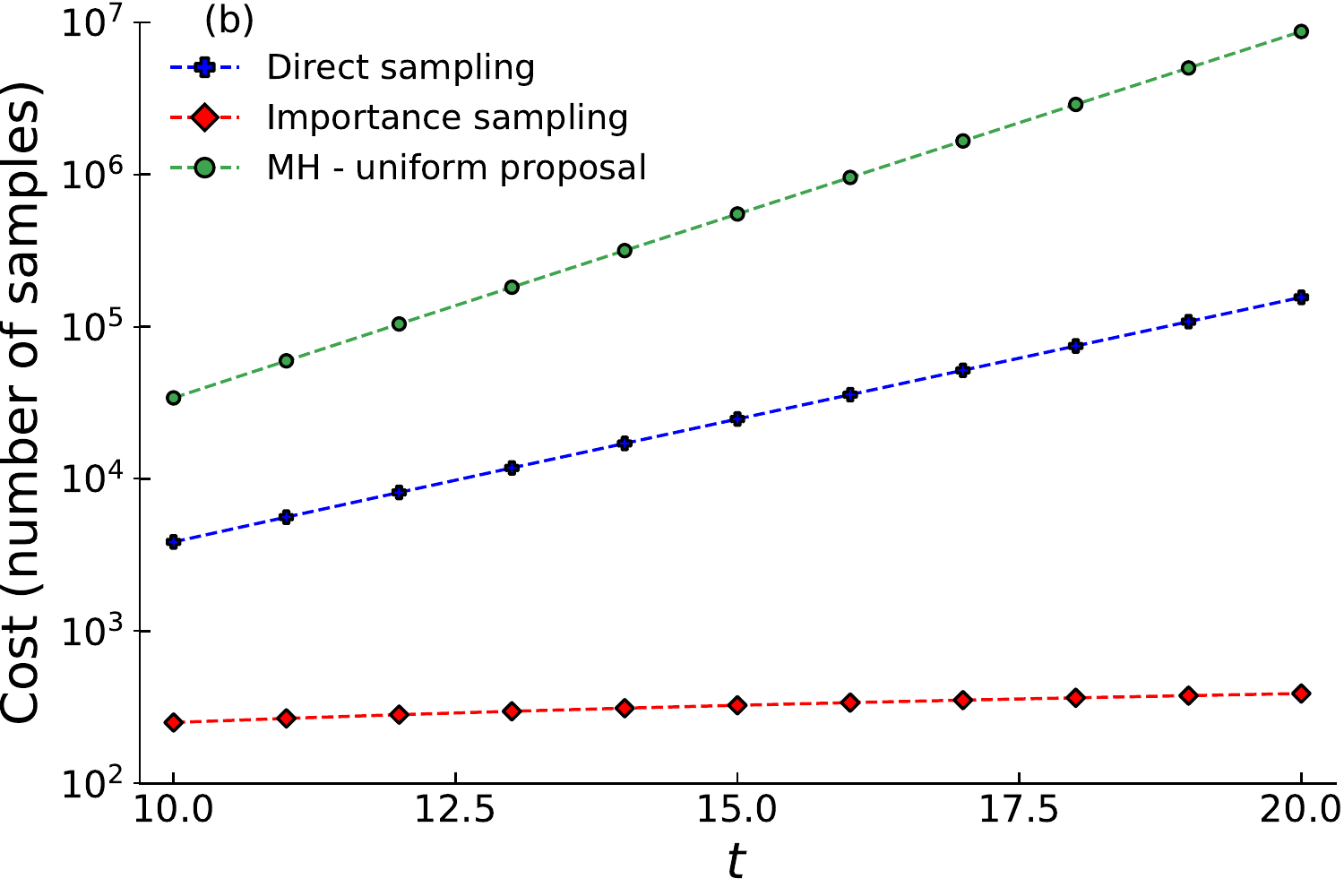}
  \label{costray}
\end{subfigure}
\caption{Improved efficiency of importance sampling in a 2D diffusive system following the distribution~\eqref{rayleigh}. (a) The Rayleigh distribution~\eqref{rayleigh} and its weighted version obtained from Eq.~\eqref{importancedistribution} with $\beta = \beta^* \approx -1.47$. (b) Computational cost of estimating the integral~\eqref{integral} with $r^*=t/2$ for $\rho(r;t)$ given by the Rayleigh distribution. The cost is measured by the number of samples needed to achieve an accuracy $e=10\%$. Three different sampling methods are considered: direct sampling~\eqref{costdirect}, importance sampling~\eqref{costimportance} with $\beta^*=-r^*/(2Dt)$, and the Metropolis-Hastings method with uniform proposal (see text). In both plots the curves are theoretical with $D = 0.17$ and $t=20$.}
\label{rays}
\end{figure*}

In the following, we compare direct sampling with importance sampling for the estimation of~\eqref{integral} when the distribution $\rho(r; t)$ is exactly known. Let us define the set $A_{t} := \{r: r > r^* \}$  for a given $t$. The direct estimator $\hat{I}^{\rm direct}_{t}$ of~\eqref{integral} for $M$ samples drawn from $\rho(r; t)$ is given by
\beq
\hat{I}^{\rm direct}_{t} = \frac{1}{M} \sum_{i=1}^M 1_{A_{t}}(r_i) \, ,
\label{directestimator}
\eeq
where $1_A$ is the characteristic function of the set $A$. This is an unbiased estimator that converges almost surely in the limit $M \to \infty$ by the Strong Law of Large Numbers~\cite{christian2004monte}. For a finite $M$, the error associated with a generic estimator $\hat{I}$ is 
\beq\label{eq.e}
e := \frac{\sigma{[\hat{I}]} }{I_{t}}
\eeq
where $\sigma^2{[\hat{I}]}$ is the variance of the estimator (and therefore $\sigma{[\hat{I}]}$ is the standard deviation). For the estimator~\eqref{directestimator} this is calculated as~\cite{robert2004monte}
\beq
\sigma^2[\hat{I}^{\rm direct}_{t}] = \frac{1}{M} \int_0^\infty [1_{A_{t}}(r) - I_{t} ]^2 \rho(r; t) \d r \, ,
\label{variance}
\eeq
and therefore the error is given by
\beq
e_{\rm direct} = \sqrt{\frac{1 - I_{t}}{M I_{t}}} 
\eeq
and the cost for a fixed error in terms of the number of samples is
\beq
M^{\rm direct}_{t}  = \frac{1 - I_{t}}{e_{\rm direct}^2 I_{t}} \, ,
\label{costdirect}
\eeq
where we put labels over $M$ to make explicit that it is associated with a sampling method and some observation time.

In the importance sampling method we draw $M$ (independent) samples from the weighted distribution~\eqref{importancedistribution}. In such a case the unbiased estimator of the integral~\eqref{integral} is
\beq
\hat{I}^{\rm importance}_{t} = \frac{Z_\beta}{M} \sum_{i=1}^M \frac{1_{A_{t}}(r_i)}{\exp(-\beta r_i)} \, ,
\label{impestimator}
\eeq
and the variance is given by
\beq
\sigma^2[\hat{I}^{\rm importance }_{t}] = \frac{1}{M} \int_0^\infty \left[ \frac{1_{A_{t}}(r) Z_\beta}{\exp(-\beta r)} -I_{t} \right]^2 \rho_\beta(r; t)  \d r \, .
\label{varestimator}
\eeq
Introducing this in Eq.~(\ref{eq.e}) and solving for $M$, as done for the direct estimator, we find the cost for a given error $e$ as
\beq
M^{\rm importance}_{t} = \dfrac{ Z_{\beta} \int_{r^*}^\infty {\rho(r; t) \exp (\beta r)} - I_{t}^2 }{ e_\mathrm{importance}^2 I_{t}^2 } \, .
\label{costimportance}
\eeq
This formula depends on the value of the parameter $\beta$. For a given $r^*$ and $t$ we use the value $\beta=\beta^*$ that minimizes the error (and variance), obtained solving $d (\sigma^2 [\hat{I}^{\rm importance }_{t}]) / {d \beta} = 0$ (and applying the second derivative test). 

Let us illustrate the discussion above for the Rayleigh distribution~\cite{krishnamoorthy2016handbook},
\beq
\rho(r; t) = \frac{r}{2Dt}\exp
\left( -\frac{r^2}{4Dt} \right),
\label{rayleigh}
\eeq
the limiting radial distribution for a two-dimensional isotropic system that exhibits normal diffusion with coefficient $D$. 
The value of $\beta^*$ that minimizes~\eqref{varestimator} is approximately 
\beq
\beta^* \approx \frac{-r^*}{2 D t},
\label{betaestimator}
\eeq
obtained by grouping the leading exponential terms in the integrand~\eqref{impestimator} and finding the minimum \footnote{This approximation has been validated numerically and has been used to calculate the importance sampling curve in figure~\ref{rays}.}. By fixing a desired precision (e.g., $e=0.1$) in our estimation of the integral~(\ref{integral}), the cost of the direct sampling method is obtained from~Eq.~(\ref{costdirect}) and of the importance-sampling case from Eq.~(\ref{costimportance}) using approximation~(\ref{betaestimator}). In fig.~\ref{rays} we show the Rayleigh distribution~\eqref{rayleigh}, its weighted version $\rho_\beta(r; t)$, and the performance of the two estimators for a representative problem. It is evident that the importance-sampling cost is lower, and scales more slowly in the relevant limit of $t \rightarrow \infty$, with respect to the direct sampling. In this example, and also in the numerical investigations discussed below, we consider problems in which $r^*$ grows linearly with $t$
\beq\label{eq.alpha}
r^*= \alpha t + c,
\eeq
with $0<\alpha\le 1$. This reduces the number of parameters and is convenient since $\beta^* \approx -\alpha/(2D)$ in Eq.~\eqref{betaestimator} is a constant and in the limit $t\rightarrow \infty$ we have increasingly hard computations, i.e. $r^*\rightarrow \infty$ and $I_{t} \rightarrow 0$ in Eq.~(\ref{integral}).

\subsection{Metropolis--Hastings algorithm}\label{MH}

Having shown the (potential) advantage of the importance sampling technique, we face the practical problem of sampling from
the weighted distribution~\eqref{importancedistribution}. The previous calculations have assumed the distributions as given and thus considering the underlying dynamics that generates them has not been needed. In the remainder of this section we introduce the Metropolis--Hastings (MH) algorithm as a way of sampling from $\rho_{\beta}(r; t)$ for a given dynamical system, and we show why a naive application of this traditional method fails to achieve the improvement in the computational cost obtained in the previous section.

Recall that $r_{t}: \Gamma \rightarrow \mathbb{R}$, and that by using the natural measure over $\Gamma$ (or a uniform distribution) we sample displacements from $\rho(r; t)$. Thus for sampling displacements from the weighted distribution~\eqref{importancedistribution} we have to change the measure over $\Gamma$. It can be shown that by constructing a Markov chain over $\Gamma$ with equilibrium distribution $\pi(x) \propto \exp \left[-\beta r_{t}(x) \right] $ , the sampling of~\eqref{importancedistribution} can be achieved; see discussion in~\cite{leitao2017importance}. The construction of this Markov chain can be done via a MH algorithm, which has the following steps

\begin{itemize}
\item \textbf{Initialize.} Choose uniformly an initial condition $x$ and the observation time $t$.
\item Measure the observable $r_{t}(x)$.
\item \textbf{Proposal.} Propose a new initial condition $x'$, drawn from a proposal distribution  $g(x \rightarrow x')$.
\item Measure the observable $r_{t}(x')$.
\item \textbf{Acceptance.} Accept the new state according to the following rule:
\beq
a(x \rightarrow x') = \min \left\{ 1, \frac{g(x' 
\rightarrow x)}{ g(x \rightarrow x')} \frac{ \pi(x') }{ \pi(x)} \right\}
\label{metropolis}
\eeq
\item If the state is accepted, set $x = x'$.
\item  Return to the proposal step.
\end{itemize}

The MH and other Markov Chain methods do not produce independent and identically-distributed (iid) random variables; rather, there is a correlation between samples which increases the cost, given by~\cite{robert2004monte}
\begin{equation}
M^{\rm{MH ,importance}}_{t} = M_{t}^{\rm importance} \kappa(x^1, x^2, \ldots)  \, , 
\label{costcorr}
\end{equation}
where $\kappa = 1 + 2 \sum_{i = 1}^\infty {\rm corr} (1_{A_{ \rm t}}(r(x^1)), 1_{A_{\rm t}}(r(x^i)) )) \ $ is the autocorrelation time associated with the sequence $\{x^i \}$, and $\mathrm{corr}$ is the autocorrelation function~\cite{thompson2010comparison}. This is relevant when the proposal distribution depends on the state $x$.

A naive  implementation of the MH algorithm uses a uniform distribution in the proposal step, i.e.
\beq
g(x \rightarrow x') = \frac{1}{|\Gamma|} \, .
\label{unifprop}
\eeq
{The problem with this proposal is that it it is difficult to sample the weighted distribution $\rho_\beta(r, t)$ for large $|\beta|$ (or, equivalently, $r^*$). This happens because typically the new state $x'$ has a value of the observable $r_{t}(x')$ distributed as $\rho(r; t)$, which is peaked at values far from the peak of $\rho_\beta(r, t)$ (c.f.~Fig.~\ref{rays}). This implies that the proposal is non-local -- the distribution of $r-r'$ is peaked at increasingly negative values -- and the acceptance~\eqref{metropolis} vanishes.  This can be calculated explicitly as follows.} Assuming that convergence to $\rho_{\beta}(r; t)$ has been achieved, the cost of the algorithm can be estimated by calculating the mean acceptance. By proposing following~\eqref{unifprop} the new states would have a displacement drawn from $\rho(r; t)$, as in the direct sampling, with the only difference that not every new state $x'$ is accepted. The acceptance is controlled by using Eq.~\eqref{metropolis}, whose mean value can be calculated as
\begin{eqnarray}\label{meana}
\bar{a}_{t} &=& \avg{a( x \rightarrow x')} =  \notag \\
 &=& \int_0^\infty d r\int_0^\infty dr' \rho(r'; t) \rho_{\beta}(r; t) ({\rm min} \left\{ 1, \exp(-\beta(r' - r)) \right\}) \, . \notag \\
\end{eqnarray}
The cost of the method is thus simply $M^{\rm direct}_{t}/ \bar{a}_{t}$ (c.f. equation~\eqref{costdirect}; note that the accepted states are iid and thus $\kappa=1$ in Eq.~\eqref{costcorr}). The results for the Rayleigh distribution can be computed and are shown as the curve \emph{MH--uniform proposal} in figure~\ref{rays}. It is evident that this method is extremely inefficient, requiring more samples to estimate~\eqref{integral} than the direct method. This happens because $\bar{a}_{t}$ goes to zero for large $t$ and $r^*$. Therefore, to be successful in exploiting the advantages of the importance sampling technique a clever proposal is required.

As a  summary of this section,  we have shown that the sampling distribution~\eqref{importancedistribution} has the potential to dramatically increase the efficiency of the sampling of the tails of a general distribution $\rho(r; t)$. However, a naive approach to generate the sampling distribution~\eqref{importancedistribution} -- based on MH and a uniform proposal $g(x \rightarrow x')$ -- is not sufficient to achieve this goal. In the next section we introduce an improved proposal distribution $g(x \rightarrow x')$ which we later show to be able to obtain the desired improvement in the scaling of the efficiency.

\begin{figure*}
\begin{subfigure}{0.5\textwidth}
  \centering
\includegraphics[width=8cm, height=6.5cm]{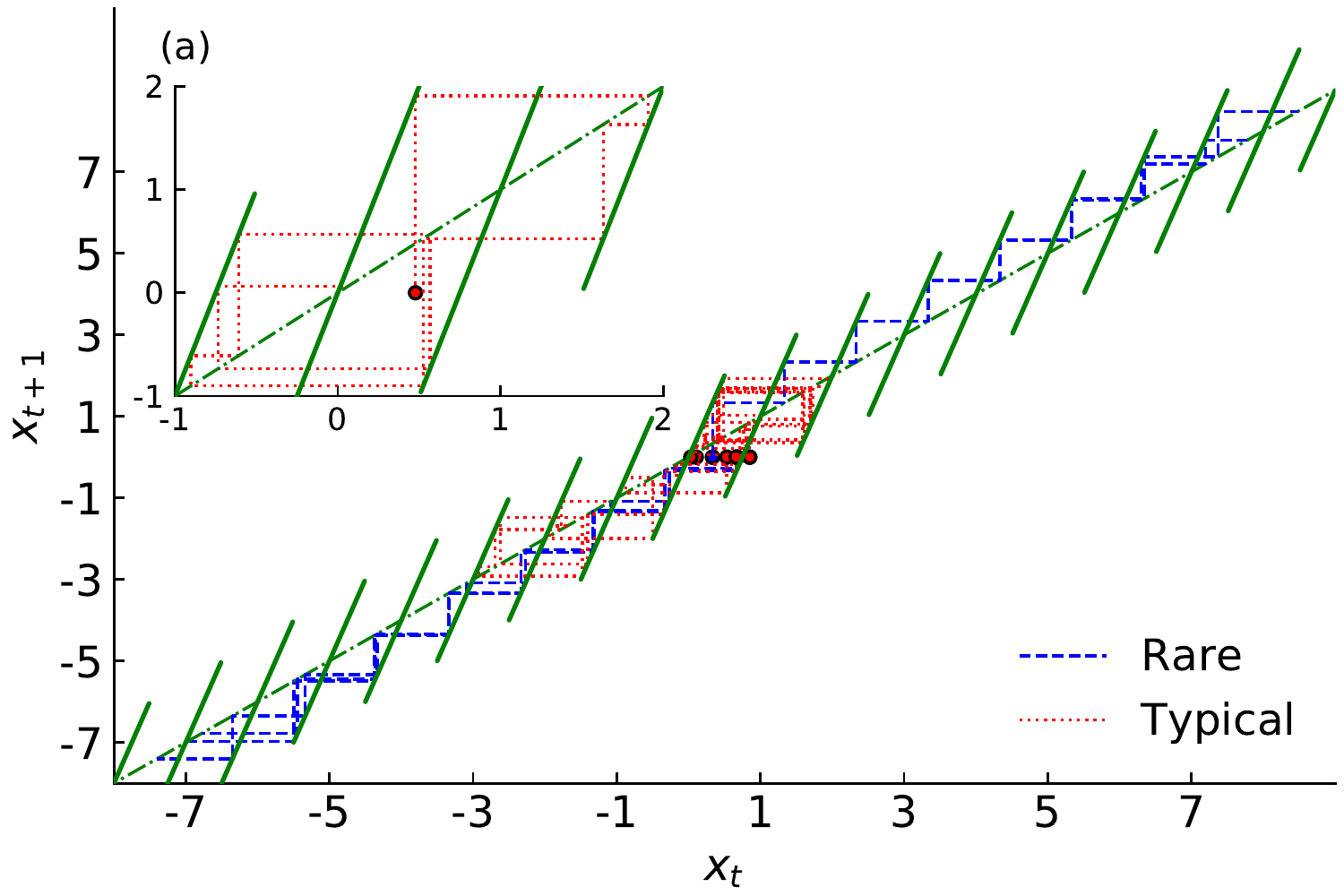}
  \label{trajecbox}
\end{subfigure}%
\begin{subfigure}{.5\textwidth}
 \includegraphics[width=8cm, height=6.5cm]{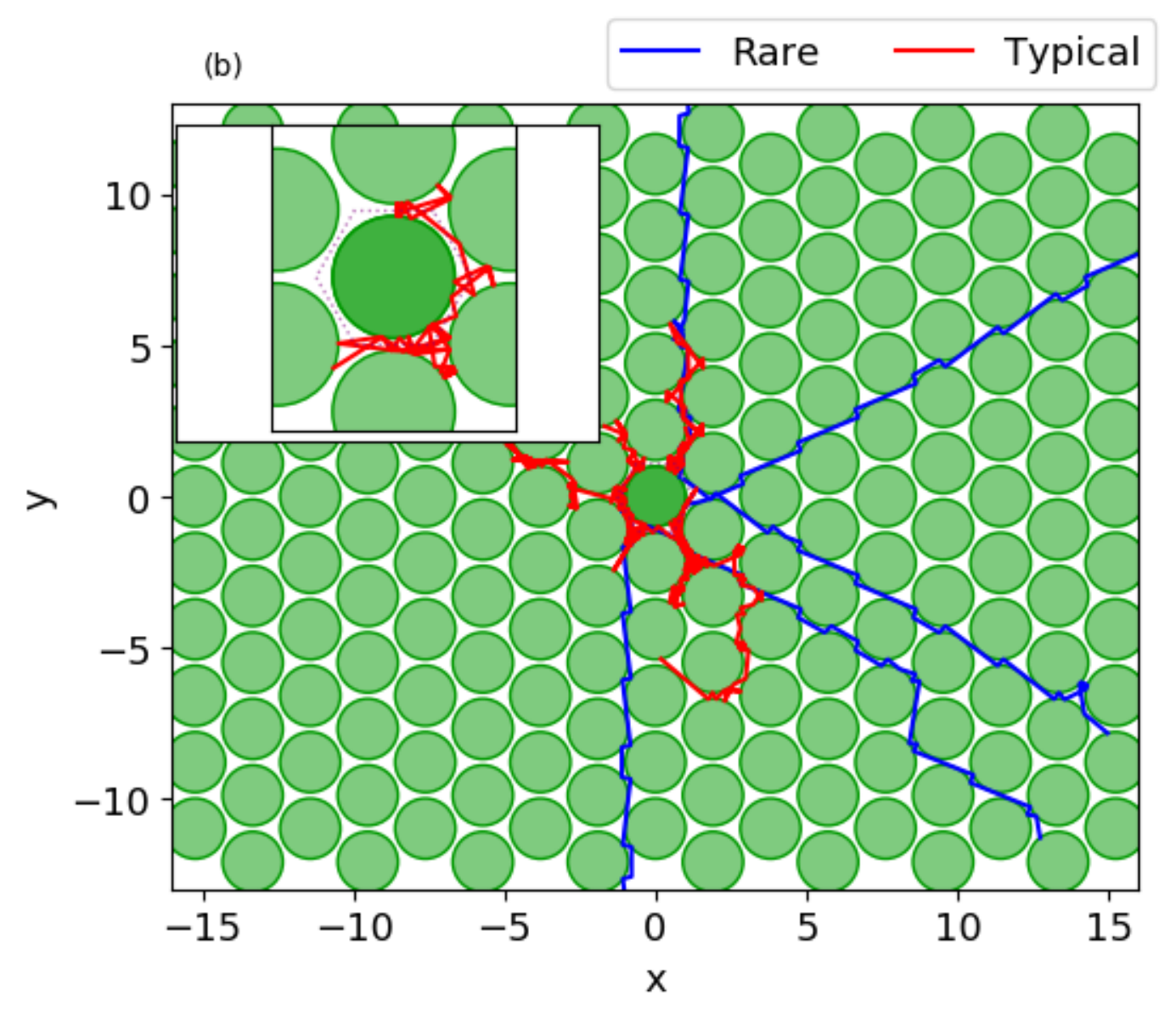}
  \label{trajeclorentz}
\end{subfigure}
\caption{Two deterministic dynamical systems showing diffusive behaviour: (a) Box map and (b) Lorentz gas. The insets (upper left corner) show a single trajectory in the fundamental cell defining the dynamics. The main panel shows the spatially extended system, composed of the periodic repetition of the fundamental cell, with five typical ($r\approx \avg{r}$, in red) and five rare ($r$ in the tails of $\rho(r;t)$, in blue) trajectories. All trajectories were initiated in the set $\Gamma = [0,1)$ for the box map and $\Gamma = S^1 \times \mathcal{D}$ for the Lorentz gas,  where $S^1$ is the unit circle (velocity space) and $\mathcal{D}$ the configuration space in a single cell.  }
\label{trajecs}
\end{figure*}

\section{Proposal distribution}
\label{algorithm}

The MH algorithm that we propose follows the same steps listed above, but constructs a proposal distribution following the general scheme set up in Ref.~\cite{leitao2017importance} (previous motivation may be found in~\cite{tailleur2007probing, geiger2010identifying, leitao2013monte, leitao2014efficiency}).
There, an explicit algorithm was constructed for sampling states in the tails of the Finite-Time-Lyapunov Exponent distribution for closed systems and the escape-time distribution for open systems. Here, we consider the displacement distribution~$\rho(r; t)$ in spatially extended systems. 

The idea is to construct an improved proposal distribution $g(x \rightarrow x')$, exploiting correlations between $x$ and $x'$ (and thus between $r_{t}(x)$ and  $r_{t}(x')$). We introduce two general proposals with a free parameter, the correlation time $t_*$, which, when adjusted correctly, achieves the goal of efficiently correlating the motion in $\Gamma$ with the landscape of $r_{t}$:

\begin{itemize}

\item[i)]  The \emph{local proposal}, consisting of drawing the next state $x'$ (in the Markov chain) at a distance $\norm{x-x'}$ drawn randomly from a half-Normal
\beq
g^{\rm local}(x \rightarrow x') = \sqrt{\frac{2}{ \pi \sigma^2(x)}} \exp\left(-\frac{\norm{x' - x}^2 }{2 \sigma^2(x)} \right) \,.
\label{localproposal}
\eeq 
 Following Ref.~\cite{leitao2017importance}, the typical length $\sigma(x)$ should be chosen as
\beq
\sigma(x) =  \Delta \exp(- \bar{\lambda} t_*(x)) \, .
\label{sigmalocal}
\eeq
where $\Delta$ a constant associated with the scale of the same order as $|\Gamma|$, $\bar{\lambda}$ is the mean value of the Lyapunov exponent, and $t^*(x)$ is the correlation time that quantifies for how long the trajectories that departs from $x$ and $x'$ are similar.  It may be the case that this proposal proposes a new state $x'$ out of the set of initial conditions $\Gamma$; in such a case the proposal step in the MH algorithm is set as $x' = x$.

\item[ii)]  The \emph{shift proposal}, consisting of proposing the next state $x'$ by integrating (or iterating) the motion  from $x$ during a certain time~{$\tilde{t}$} (that depends on the value of $r_{t}(x)$) and then taking the resulting state (or its projection into $\Gamma$), combining with equal probability a forward and a backward motion:\beq
g^{\rm shift}(x \rightarrow x') = \frac{1}{2} \delta(x' -  \phi^{\tilde{t}}(x) ) + 
 \frac{1}{2} \delta(x' -  \phi^{-\tilde{t}}(x) ) \, ,
\label{shiftproposal}
\eeq
with $\tilde{t}$ drawn from an arbitrary  distribution $f(\tilde{t})$ with mean $t_\mathrm{shift}(x) = t - t_*(x)$ and support $[0,t]$. In comparison to Ref.~\cite{leitao2017importance}, this is a more general way of setting the shift proposal, which is essential to guarantee the detailed balance condition of the MH algorithm\footnote{The ratio of proposals -- appearing in the acceptance probability in Eq.~(\ref{metropolis}) -- is given by
\beq
\frac{g^{\rm shift}(x' \rightarrow x)}{ g^{\rm shift}(x \rightarrow x')}  = \frac{ f(\tilde{t}; t_{\rm shift}(x'))} {f(-\tilde{t}; t_{\rm shift}(x))} \, .
\eeq
The
  choice of a simple function $\tilde{t}=t_\mathrm{shift}(x)$ (i.e., $f(\tilde{t};t_{\rm shift})=\delta (\tilde{t}-t_{\rm shift}(x))$), as  proposed in Ref.~\cite {leitao2017importance}, typically leads to a vanishing  ratio and thus to a vanishing acceptance (except in the trivial case of a constant shift, i.e. $t(x)= \rm{cst}$, independent of $x$).}. In our simulations, $f(\tilde{t})$ is a truncated normal distribution. 
\end{itemize}

In both types of proposal summarized above the main piece of the puzzle is to find the formula that relates adequately the correlation time $t_*(x)$ with the observable $r_{t}$. We deduce it by departing from the same premise as any efficient MH algorithm, which is to maintain the acceptance ratio away from  0 and 1. By using the fact that correlated trajectories are statistically indistinguishable up to time $t_*$ and that for a diffusive system the mean square displacement scales linearly with time (c.f. Eq.~\eqref{coefdif}) we obtain (see appendix~\ref{formula}) 
\beq
t_* (x)= \max \Bigl\{ t_{\rm min}, t - \bigg \lvert \frac{a-1}{2D \beta (1 -\frac{r^2}{2Dt} ) } \bigg \rvert \Bigr\} \,,
\label{maintstar}
\eeq
where $D$ is the diffusion coefficient, $r=r_{t}(x)$ is the displacement of the initial condition $x$, $0<a<1$ is a constant, and $t_{\rm min}$ is a parameter that accounts for the fact that the correlation time may assume a minimum value (e.g., $t_{\rm min}>0$)~\footnote{In the simulations we use $a = 0.5$ and $t_{\rm min} = 9/10 t$. The latter choice corrects a systematic deviation that occurs around $\frac{r^2}{2Dt} = 1$ in the estimation of $\rho(r; t)$.}. 

The total proposal rule combines the two proposal rules discussed above as
\beq
g^{\rm s+l}(x \rightarrow x') = \frac{1}{2} g^{\rm shift}(x \rightarrow x') + \frac{1}{2} g^{\rm local}(x \rightarrow x') \, ,
\label{ourprop}
\eeq
with the adjustable parameter $t_*(x)$ controlled by~\eqref{maintstar}.

In the theoretical results above the Lyapunov exponent~$\lambda$ and diffusion coefficient~$D$ of the system appear, in Eqs.~(\ref{sigmalocal}) and~(\ref{maintstar}) respectively. In a practical implementation of the method in cases in which these quantities are unknown, the simplest way to proceed is to estimate them from a standard direct-sampling simulation -- both quantities rely on typical trajectories which are not computationally demanding to estimate. In the problem of transient chaos, Ref.~\cite{leitao2013monte} showed that using the correct of $\lambda$ is essential and introduced an adaptive proposal which converges to that value (see also Ref.~\cite{leitao2017importance} for further discussions, e.g., on alternatives based on the finite-time Lyapunov exponent). For the case of $D$, below we show numerical evidence that a rough estimation is sufficient, i.e., our method remains efficient even if the value of $D$ used in Eq.~(\ref{maintstar}) differs from the true diffusion coefficient of the system.

\begin{figure*}
\begin{subfigure}{0.32\textwidth}
  \centering
 \includegraphics[width= \linewidth]{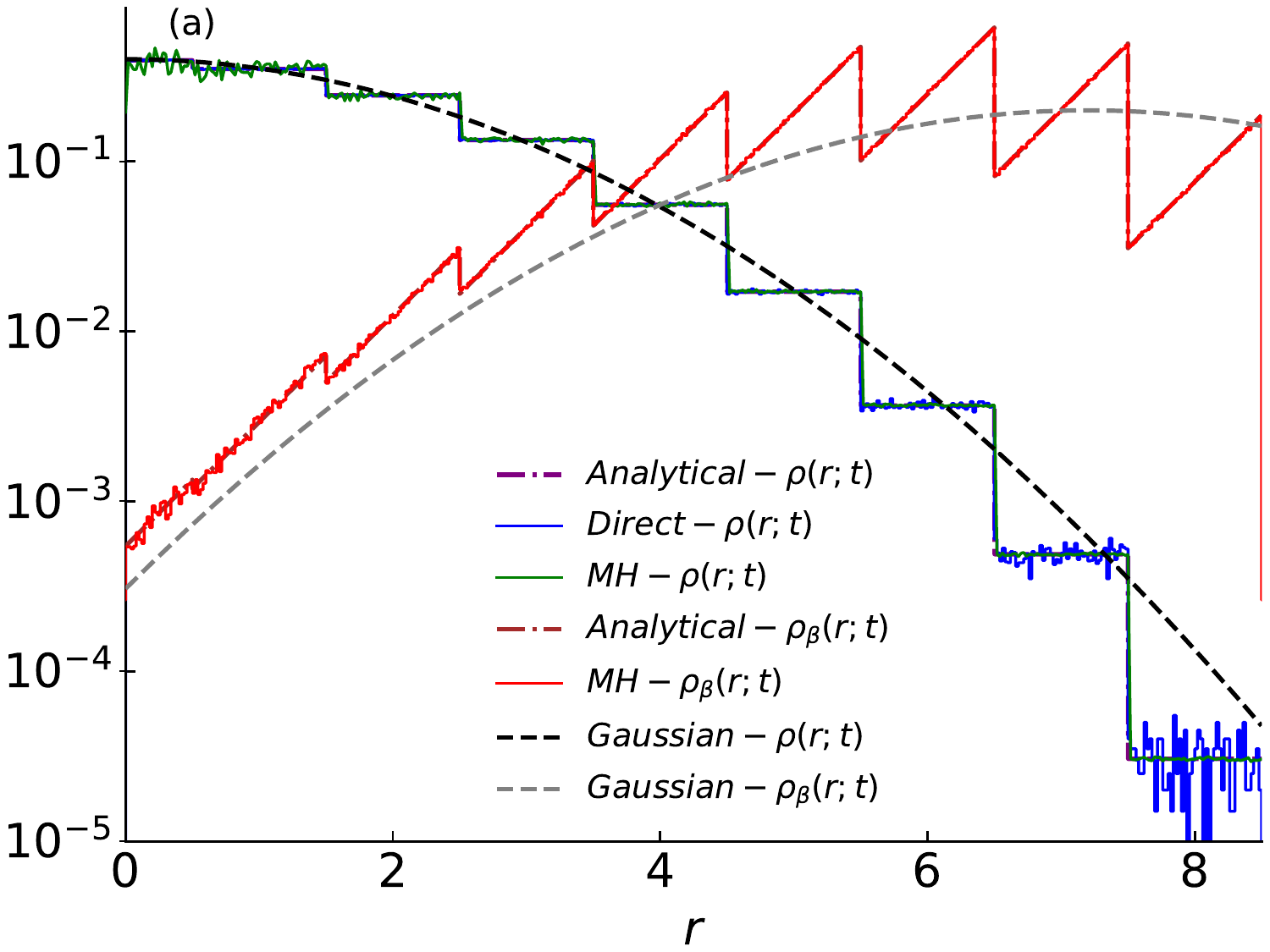}
\label{rhos}
\end{subfigure}\hfill
\begin{subfigure}{.32\textwidth}
 \includegraphics[width= \linewidth]{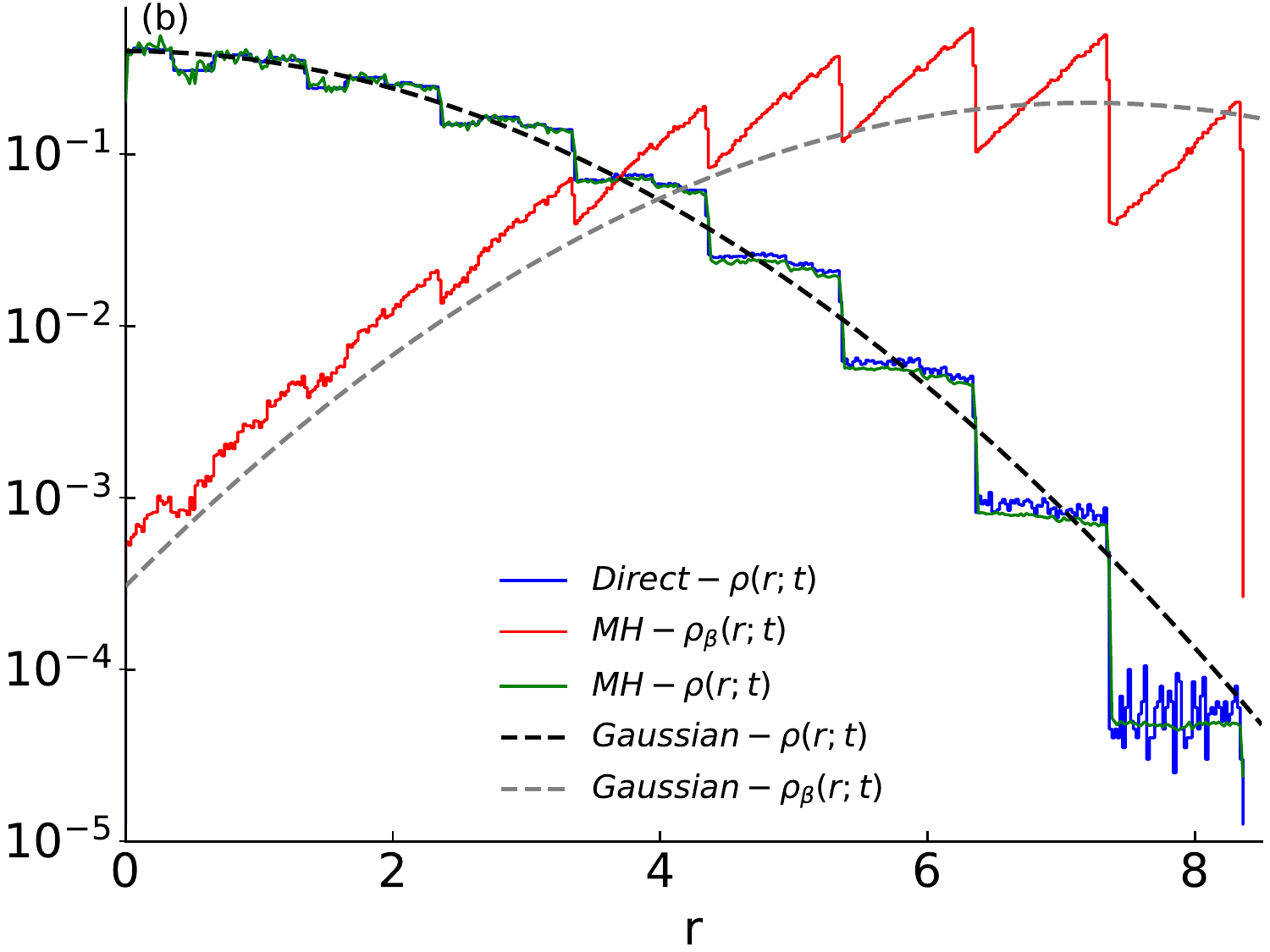}
\label{newbox}
\end{subfigure}\hfill
\begin{subfigure}{.32\textwidth}
\includegraphics[width= \linewidth]{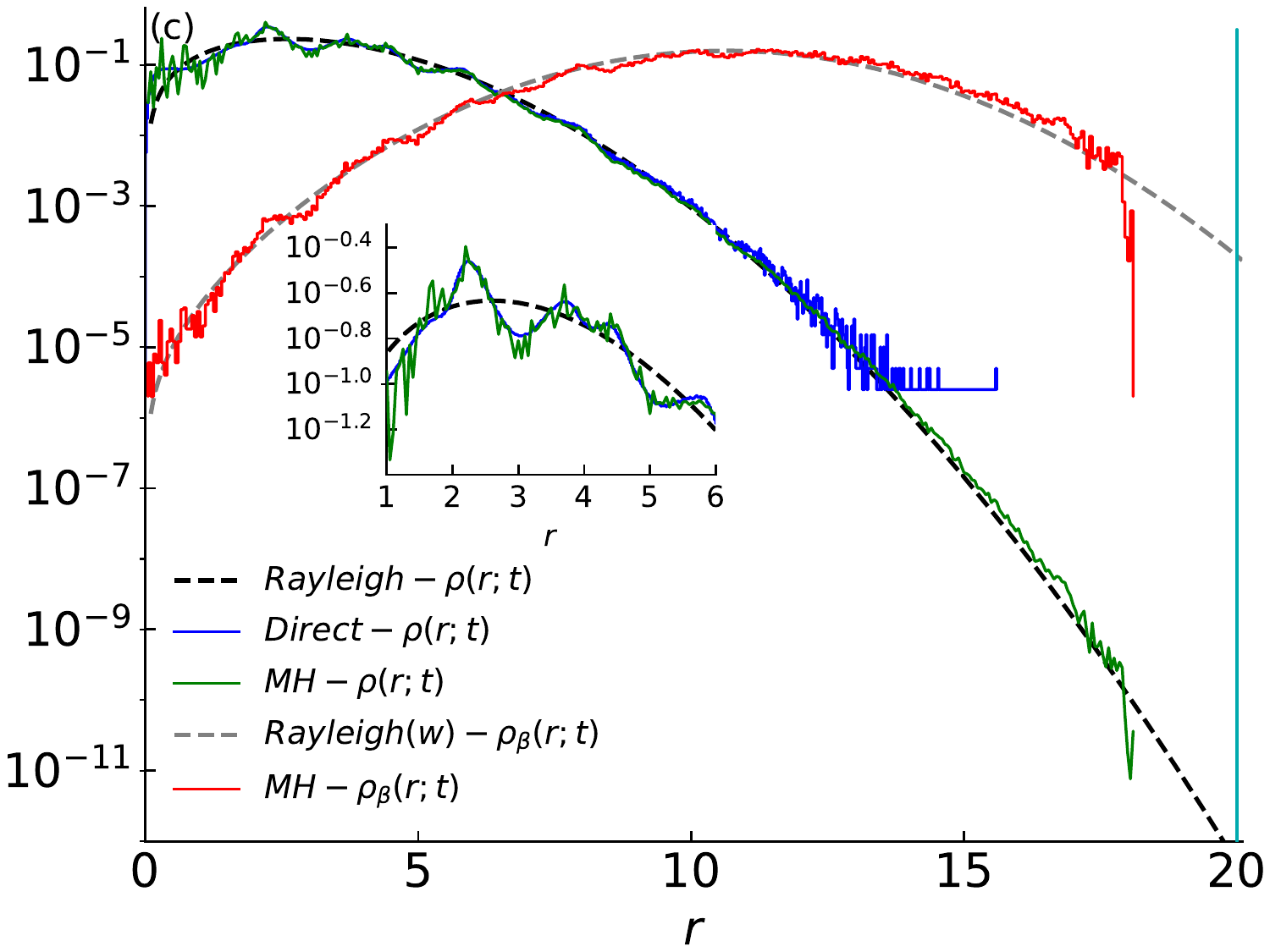}
  \label{lorentzR}
\end{subfigure}

\caption{Estimation of $\rho(r;t)$ and $\rho_{\beta}(r; t)$ using the proposed importance-sampling method (MH in red). Results are compared against the direct-sampling method (Direct in blue), against the diffusive approximation (dashed line, Gaussian and Rayleigh distributions), and the exact result~\eqref{binomial} for the case of the box map with integer $a$ (dashdotted line). Results for the two numerical methods are based on {$M=10^7$} samples and the estimation of $\rho(r;t)$ based on the importance-sampling method (in green) is obtained by reweighting $\rho_{\beta}(r;t)$~\cite{landau2014guide}.  (a) Box map, with parameters  $a = 4$, $D = 1/4$, $\beta = -1.8$ and $t = 8$.  (b) Box map, with parameters $a = 3.7$, $D = 1/4$, $\beta = -1.8$ and $t = 8$. (c) Lorentz gas, with parameters  $\beta = -1/4$, $D = 0.17$ and $t = 20$. Inset: magnification around the peak of the distribution showing strong deviations from the Rayleigh distribution. }
\label{lorentz}
\end{figure*}

\section{Numerical tests}

In this section, we test the validity of our method via numerical simulations in two systems: 

\paragraph{The box map} is a one-dimensional piecewise-linear map $x_{t+1}=M_a(x_t)$ -- defined in Eq.~(\ref{eq.map}) below -- that is one of the simplest chaotic system that exhibits normal diffusion\cite{fujisaka1982chaos}. The parameter $a$ controls the chaoticity (for $|a| > 1$ the system is chaotic with Lyapunov exponent $\lambda := \ln a$); the diffusion coefficient $D$, shows a fractal dependence on $a$, with analytical formulas known for $a \geq 2$ when $a$ is an integer~\cite{klages1995simple, klages2007microscopic}. Here we use mostly $a = 4$, in which case the distribution of $x_n$ is a binomial distribution (see appendix~\ref{boxmap} for details), and thus the convergence to a Gaussian in the limit that $t \rightarrow \infty$ is expected (diffusive regime). 
 
\paragraph{The Lorentz gas} is one of the most studied deterministic chaotic systems. In the version considered here, it is a periodic array of circular scatterers that forms a triangular lattice~\cite{dettmann2014diffusion}. Depending on the separation between the scatterers the following two regimes are defined: finite horizon and infinite horizon~\cite{gaspard1995chaotic}.  In the first regime, which we consider here, the free motion of the particle is bounded, and ensembles exhibit normal diffusion, with a central limit theorem having been proved in the limit $t \rightarrow \infty$~\cite{bunimovich1981statistical} (the sequence  $\vec{r}(t)/ \sqrt{t}$ converges weakly, or in distribution, to a normally distributed random variable. This asymptotic distribution, in terms of the scalar $r$, is the Rayleigh distribution~\eqref{rayleigh}.).  For a finite time $t$, however, the distribution is unknown. We consider a triangular unit cell with a centered disk of unit radius and with a lattice space between the centers of the disks of $d = 2.2$, following the notation of~\cite{gaspard1995chaotic}. We calculate the mean Lyapunov exponent following~\cite{dellago1996lyapunov} and obtain $\bar{\lambda} = 2.10$~\cite{Datseris2017}. In addition, we take the diffusion coefficient equal to $D =  0.17$, as reported in~\cite{gaspard1995chaotic}.  

These two dynamical systems are illustrated in Fig.~\ref{trajecs}, together with a few trajectories with typical and rare spreading in space. This shows how typical and rare trajectories fundamentally differ from each other, illustrating also the  significance of our algorithm not only in estimating $\rho(r;t)$,  but also in identifying actual trajectories in the tail of this distribution. Next we compare, for each of these two systems, our sampling method with the direct sampling and the MH algorithm with alternative proposals.  The code used in our simulations is available in Ref.~\cite{codemcmc}.

\subsection{Sampling of $\rho_\beta(r; t)$}

The first test of our method is to confirm that it correctly samples from $\rho_\beta(r;t)$ for a given $\beta$. The results shown in Fig.~\ref{lorentz} confirm that this is the case, leading to an improved statistical accuracy in the estimation of the tails of $\rho(r;t)$ (compare the green and blue curves in the figure). Another successful test of our method is that it is able to correctly estimate the deviations from the diffusive approximation (e.g., due to finite $t$). These deviations appear both at the core of the distribution -- e.g., $r\approx 3$ in the Lorentz gas shown in panel (c) -- and at its tail. While these deviation are known exactly for the case of the box map with $a=4$, serving as an additional test to our method, the shape of $\rho(r;t)$ in the Lorentz gas is not known for finite $t$ so that we rely solely on the direct-sampling and MH numerical estimations.  (The systematic deviations at the core of the distribution for long times were studied by one of the present authors in~\cite{sanders2005fine}.) Both these estimations are in agreement on the deviations in the core of the distribution, but only the MH method is able to capture the behavior in the tail. The existence of deviations in the tail can be intuitively understood by realizing that for any $t$ there is a maximum travel distance $r^\dagger$, while the Rayleigh distribution has support in $[0,\infty)$. The constant velocity $|v|=1$ of trajectories in the Lorentz gas implies that $r^\dagger \le t$; in the finite-horizon regime considered here, the inequality is actually strict, $r^\dagger < t$. Numerical estimations of $r^\dagger$ and $\rho(r;t)$ for $r\lesssim r^\dagger$ are difficult to obtain using the direct-sampling method; for example, in the example of Fig.~\ref{lorentz}(c) the Rayleigh distribution in the interval $r \in [18,20]$ has a measure of $\approx 10^{-11}$, so that in principle we expect that of the order of $10^{11}$ directly-sampled initial conditions would be needed to sample one trajectory in this interval. In contrast, our importance-sampling method reduces the cost of these estimation by biasing the samples towards large values of $r$; e.g., in the case shown in the figure we estimate $r^\dagger \approx 18.08$ and $\rho(r^\dagger) \approx 3.6 \times 10^{-11}$ based on solely {$10^7$ samples}.

As an additional test of our method, we compare the direct and MH sampling for the box map with $a=3.7$, where the exact distribution $\rho(r; t)$ is unknown (Fig.~\ref{lorentz}(b)). The exact value of $D$ for this case is also unknown and we use in Eq.~(\ref{maintstar}) the value of $D$ for $a=4$. Our method (MH, green line) agrees with the direct sampling (Direct, blue line) in the description of the fine structures of $\rho(r;t)$ and, additionally, provides an increased statistical accuracy in the estimation of the tails. This success of our method shows also that it is robust with respect to the value used for $D$ in Eq.~(\ref{maintstar}).

\subsection{Acceptance}

\begin{figure}[h!]
\begin{subfigure}[b]{0.45\textwidth}
  \centering
  \includegraphics[width=1\linewidth]{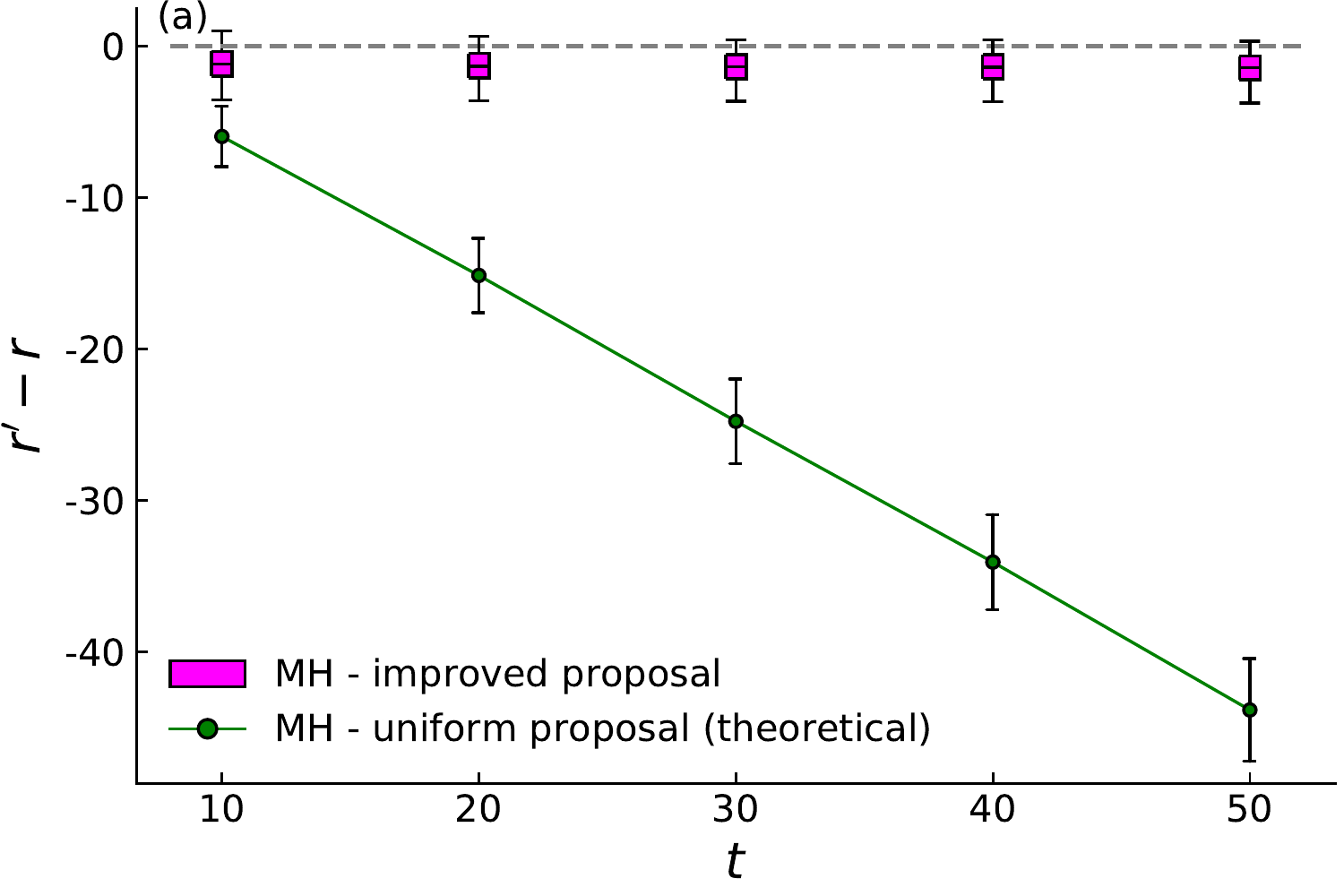}
   \label{box}
\end{subfigure}

\begin{subfigure}{0.45\textwidth}
  \centering
 \includegraphics[width=1\linewidth]{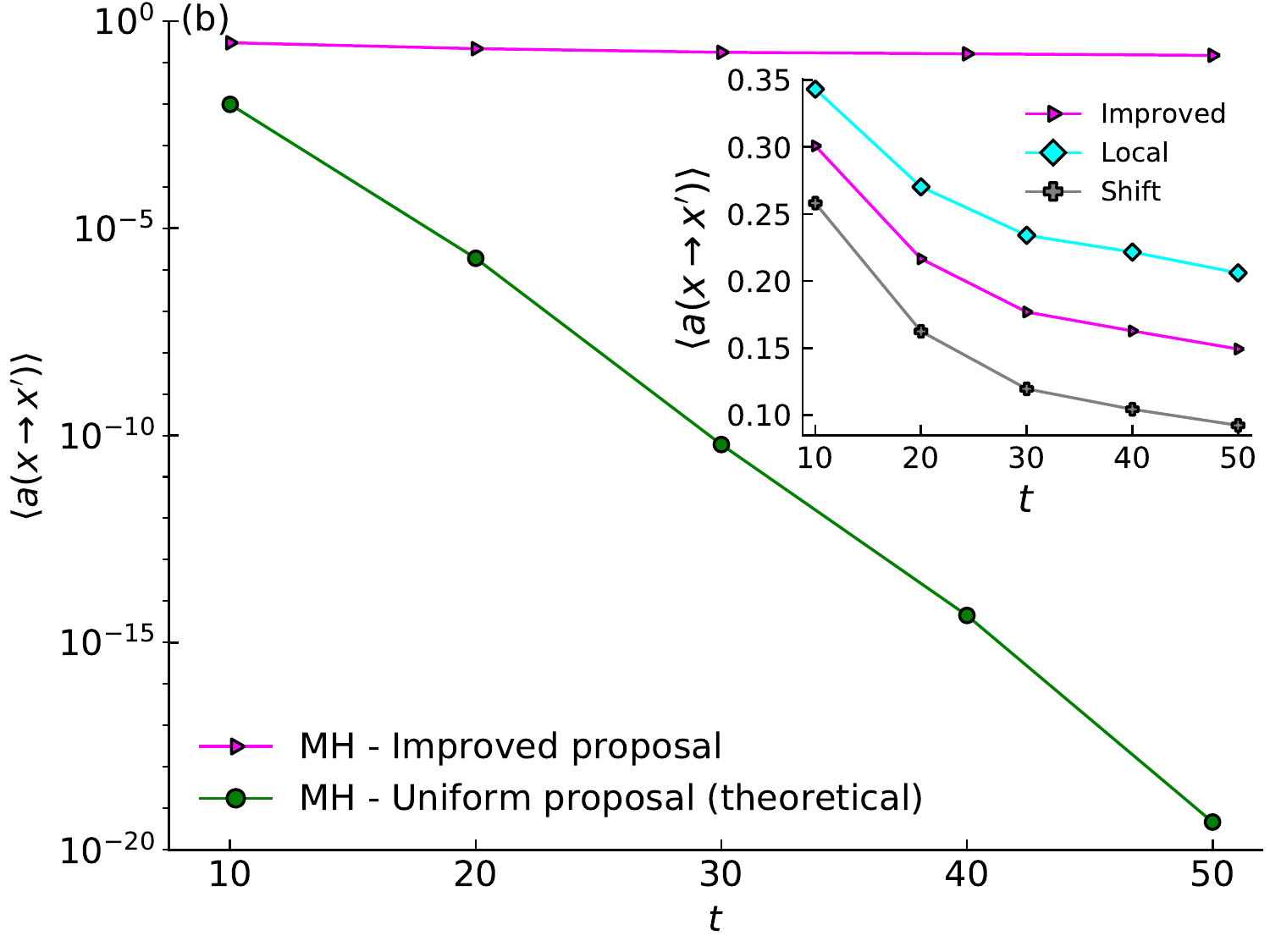}
  \label{acc1}
  \end{subfigure}
\caption{
  Comparison of the (naive) uniform proposal~\eqref{unifprop} and our proposal~\eqref{ourprop} for the box map as a function of the observation time $t$. For each $t$ we choose  $\beta=\beta^*$ that minimizes the cost in Eq.~\eqref{costimportance} for $r^* = t - 4.5$ (obtained numerically). (a)  Distribution of the locality ($r' -r$). For the improved proposal, the boxplots show the $5\%, 25\%,50\%,75\%,$ and $95\%$ quantile of the distribution (obtained numerically). For the uniform proposal, the average and standard deviation are shown (obtained analytically  as discussed after Eq.~\eqref{unifprop}, considering  $\rho(r; t)$ and  $\rho_\beta^*(r'; t)$ as the theoretical distributions for $r(x)$ and $r(x')$ respectively). (b) Mean acceptance, computed numerically for the improved proposal and using Eq.~\eqref{meana} for the uniform proposal. Inset:  same curve of the improved proposal  zoomed and  separated into its components (shift and local). In both plots, the numerical curves for the improved proposal were obtained for $10^5$ different points sampled from $\rho_\beta^*(r; t)$. }
\label{acces}
\end{figure}

The second test of our method is to check that method-specific characteristics are behaving as theoretically expected. This is an important test because, in principle, the sampling of the distribution~\eqref{importancedistribution} using the Metropolis--Hastings algorithm could work for different proposals $g(x \rightarrow x')$ and it is unclear whether the elaborate ideas we include in our method are indeed necessary. 
In Fig.~\ref{acces} we show numerical results in the Box map comparing how the acceptance (panel (a)) and locality (panel (b)) of our proposal~\eqref{ourprop} compares with the naive uniform proposal~\eqref{unifprop} for increasing $t$ ($r^*$).
The results show that while the (naive) uniform proposal becomes non-local with a vanishing acceptance for large $t$, the MH with our improved proposal preserves locality: for all $t$, on average at least $5\%$ of the proposals have positive $r'-r$ (panel (a)) and the acceptance ratio remains bounded away from zero (panel (b)). 
These results confirm our theoretical findings -- end of Sec.~\ref{MH} for the naive proposal and Sec.~\ref{algorithm} for our method -- confirming the success of our algorithm in constructing an {\it efficient} Markov Chain that converge to the weighted distribution~$\rho_\beta(r;t)$. 

\begin{figure*}
\begin{subfigure}{0.5\textwidth}
  \centering
\includegraphics[width=\linewidth]{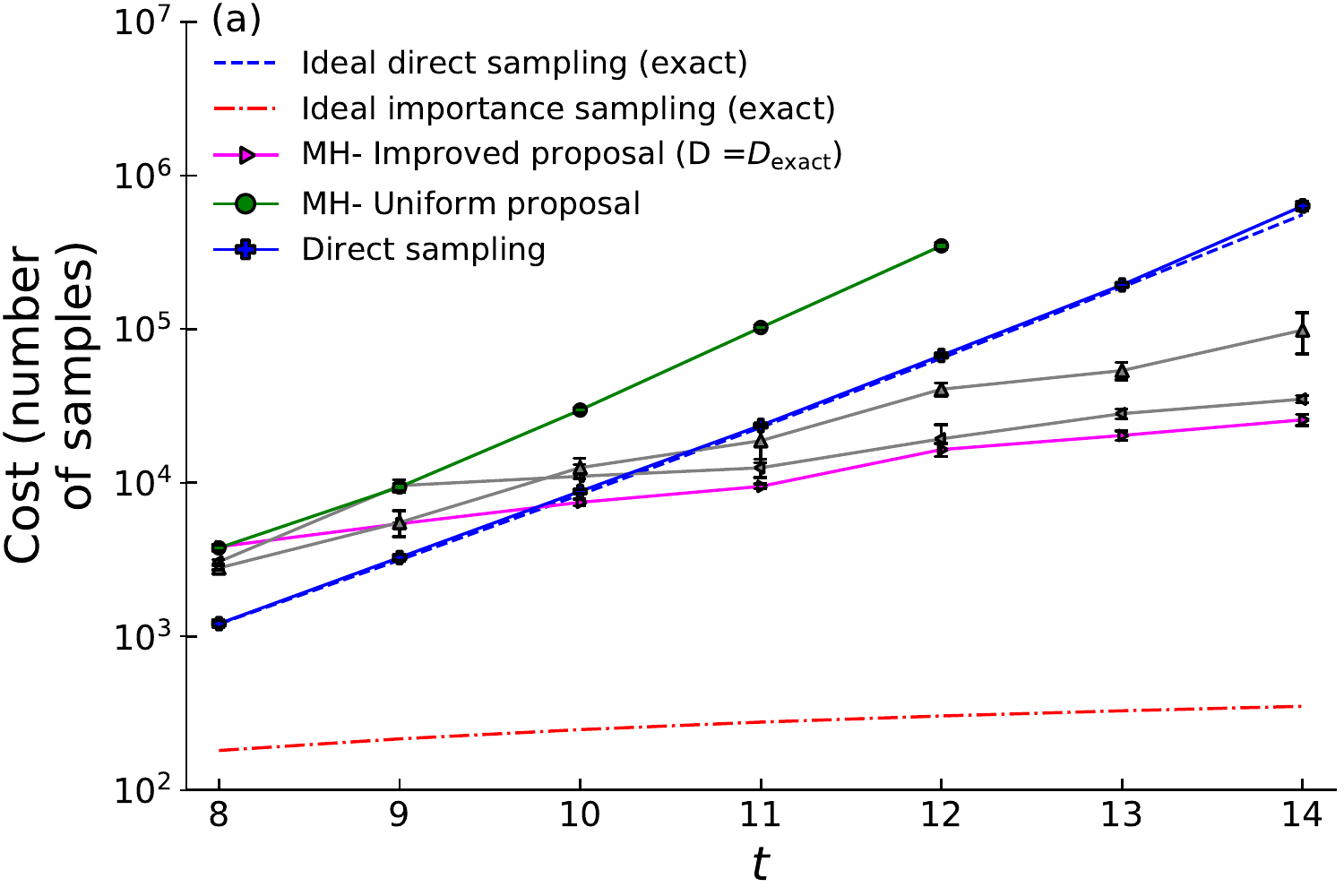}
  \label{costbox}
\end{subfigure}%
\begin{subfigure}{.5\textwidth}
 \includegraphics[width=\linewidth]{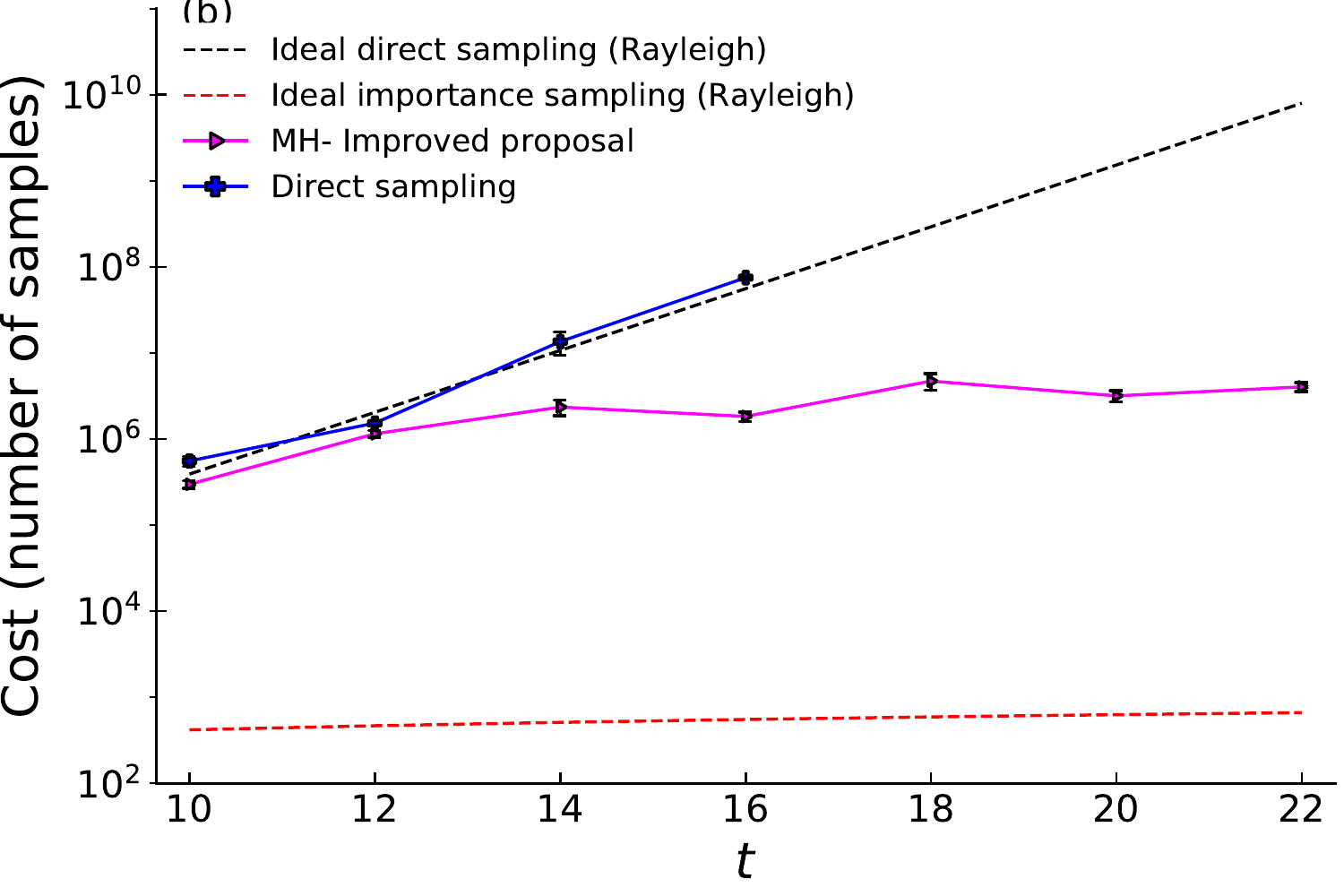}
  \label{costlorentz}
\end{subfigure}
\caption{Improved efficiency of the importance-sampling method. The computational cost of different sampling methods is measured as the number of samples needed to estimate the integral~\eqref{integral} with a an accuracy $e=10\%$. (a) Box map with $r^* = t - 4.5$, and $\beta^*$ obtained numerically by  minimizing~\eqref{costimportance}; comparison between our proposal (MH- improved proposal), the uniform proposal (MH-uniform proposal), direct sampling (numerical and theoretical results), and the ideal importance sampling (obtained introducing the exact $\rho(r; t)$, Eqs.~\eqref{binomial}, in \eqref{importancedistribution}). In addition, our proposal is implemented with two different values of $D$ in equation~\eqref{maintstar}(gray curves), namely $D = 2/3 D_{\rm exact}$ and $D = 4/3 D_{\rm exact}$.(b) Lorentz gas. We estimate the integral~\eqref{integral} with $r^* = 3t/4$ using $\beta^* = -3/(8D) \approx -2.2$ obtained from~\eqref{betaestimator}. For comparison, the theoretical results on the Rayleigh distribution~\eqref{rayleigh} are shown. To numerically estimate the cost at a given $t$, $i=1,2 \ldots, 10$ independent realizations of the numerical method are repeated and for each realization the value~$I$ of the integral~\eqref{integral} is estimated as $\hat{I}_i(M)$ for increasing number of samples $M$. The reported cost correspond to the minimum value of $M$ at which the standard deviation of the 10 estimations is $\sigma(\hat{I}_i(M))< e I$ with $e=10\%$, see Eq.~(\ref{eq.e}). The error bars is the standard error obtained repeating this whole procedure for further $10$ times. }
\label{costs}
\end{figure*}

\subsection{Efficiency}

The third and final test of our method is to compare the efficiency of our method with that of the alternatives. As a cost to measure efficiency we use the number of samples needed to obtain a predefined accuracy in the estimation of a quantity of interest, the probability of the tail of the distribution in Eq.~\eqref{integral}. Figure~\ref{costs} compares how this cost scales for increasingly hard problems (obtained increasing $t$ and thus $r^*$). It is clear that for large $t$ our method outperforms the direct sampling and naive MH method, showing not only a lower cost but, more importantly, a different scaling with $t$. 
We also compare this scaling to the ideal MH method obtained from an iid sample of $\rho_\beta(r;t)$, this is known exactly in the case of the box map (panel (a)) and is based on the Rayleigh approximation in the Lorentz gas (panel (b)). In comparison to this ideal case, the cost of our method is shifted upwards because of the unavoidable correlation of the points of the Markov Chain (c.f.~equation~\ref{costcorr}). However, the scaling with $t$ of our method seems to be the same as in the ideal case (even when $D$ is not exact), suggesting that the correlation time of the chain is not increasing with $t$.

\section{Conclusions and future directions}

In this work we presented a new algorithm based on the Metropolis--Hastings (MH) method to sample the tails of the displacement distribution~$\rho(r;t)$ in chaotic systems that exhibit normal diffusion. The core of the algorithm is  the proposal~\eqref{ourprop}, which combines local and non--local steps along the set of initial conditions. We showed that the method is more efficient than both traditional direct-sampling methods and MH methods based on alternative (uniform) proposals. The superiority of our proposal resides in the idea of introducing a correlation time $t^*$ between trajectories, based on the general framework to construct importance-sampling methods in deterministic dynamical systems developed in Ref.~\cite{leitao2017importance}. From this point of view, the main results of this work are: (i) the application of the general framework to a new class of problems (computation of transport properties in spatially extended systems),  deducing the relation~\eqref{maintstar} between $r$ and $t^*$; and (ii) a modification of the procedure to perform the shift proposal, Eq.~\eqref{shiftproposal}, that guarantees the detailed balance condition of the MH method (this modification should be considered also in other applications of the MH method in deterministic systems). The success of our method opens the possibility of identifying trajectories (and their probability) with anomalously large displacement. 

Future extensions of our method could consider cases where the chaoticity condition is broken but in which normal diffusion is still present (such as polygonal billiard channels; see, e.g., Ref.~\cite{sanders2006occurrence}) and also systems that exhibit anomalous diffusion, such as the Lorentz gas with infinite horizon~\cite{cristadoro2014measuring} and the extended Pomeau-Manneville map~\cite{korabel2007fractal}, in which efficient computational methods are required for adequately sampling the tails. We believe that our results will contribute towards the development of a new set of techniques for the computational study of transport properties in dynamical systems.

\section*{Acknowledgments}

DT acknowledges financial support from CONACYT (CVU No. 442828) and from The University of Sydney for the realization of this work. We thank J.C. Leit\~ao for his careful reading of the manuscript.  DPS thanks T. Gilbert and B. Mognetti for discussions at the beginning of this work. Financial support from CONACYT-Mexico grant 209458 and DGAPA-UNAM PAPIIT grant IN117117 are acknowledged. 

\appendix
\section{Box map}
\label{boxmap}

The so-called box map~\cite{klages1995simple, klages2007microscopic} is given by
$$ M_a: \mathbb{R} \rightarrow \mathbb{R} , \qquad  x_{t+1} = M_a(x_t) $$
with
\beq 
M_a(x) = \left\{ 
\begin{array}{c}\label{eq.map}
ax \quad,  \qquad \qquad 0 < x \leq 1/2 \\
ax + 1 - a \quad , \quad  1/2 < x \leq 1
\end{array}
\right.
\eeq
and the conditions
\begin{align*}
M_a(x + 1) &= M_a(x) + 1 \\
M_a(x) &= -M_a(-x) \, .
\end{align*}
This system is diffusive for values of the parameter $a \geq 2$. The diffusion coefficient~\eqref{coefdif} for integer values of the parameter $a$ is~\cite{klages1995simple}
$$
D(a) = 
 \left\{ \begin{array}{c}
  ((a-1)(a-2)/24 \quad, \qquad a \text{ is even}  \\
(a^2 - 1)/24 \quad, \qquad a \text{is odd}
\end{array}
\right.
$$

For $a = 4$ -- used in this paper -- the distribution of $r(t) := |x_{t} - 1/2|$ is~\footnote{The displacement is more precisely ${r:= |x_{t} - x_0|}$ but for preserving the analytical knowledge of the distribution we focus on this variable.}
\beq 
\rho(r; t) = 2 \binom{2t}{ \lfloor r + 1/2 + t  \rfloor}  \left( \frac{1}{2} \right)^{2t} \, .
\label{binomial}
\eeq
where the first part denotes the binomial coefficient and $\lfloor r + 1/2 + t  \rfloor$ is the integer part of $r + 1/2 + t$. 

\section{Derivation of main formula \eqref{maintstar}}
\label{formula}

Here we deduce the main formula~\eqref{maintstar}. This follows the main lines of the general procedure stated in~\cite{leitao2017importance}. It is remarkable that we explicitly use the fact that the system is diffusive to arrive at a simple formula.

The ideal proposal should make the \rm{average} acceptance ratio constant and independent of the initial point $x$, i.e.
\beq
\E[a(x \rightarrow x') | x] = \int_{\Gamma} a(x \rightarrow x') g(x \rightarrow x') \d x' = {\rm constant} \, .
\eeq
Setting up a theoretical framework for reaching this objective is still a challenge for chaotic systems. However, this condition may be relaxed to one that makes the average ratio between the distributions $\pi(x')/\pi(x)$ constant (c.f. Eq.~\eqref{metropolis}). To first order, for the case where $\pi(x) \sim \exp(-\beta E(x))$, this condition is rewritten as
\beq
\E \left[  {E(x) - E(x')}   | x \right] =  \frac{1 - a}{\beta}  \, ,
\label{constraint}
\eeq
where $0 < a < 1$ is a constant. In what follows we take the natural variable appearing on a diffusive process as $E(x) = r^2(x,t) \equiv r^2_{t}(x)$ and using~\eqref{constraint} we deduce a relation between this observable and $t^*(x)$.

In arbitrary dimensions, given a point $x \in \Gamma$, the displacement vector $\r$ at time $t$ can be decomposed as the vector sum:
\beq
\r(x, t) = \r(x, t_*) + \r(\phi^{t_*}(x), t - t_*) \, ,
\label{vector}
\eeq
where $t_*$ is any intermediate time, i.e. $t_* \in (0,t)$ and $\phi$ the flow (map) induced by the dynamics. The logic behind the following deduction is to use the hypothesis that for two initial conditions $x, x' \in \Gamma$ related via some proposal, their trajectories are correlated during a time $t_*$, and so the observable is correlated during this time. After $t_*$ the trajectories are independent and are random realizations of the diffusive process. 

From~\eqref{vector}, the distance $r = \norm{\r}$ satisfies for a given initial condition $x$ the following relation:
\begin{widetext}
\beq
r^2(x, t) = \r(x, t) \cdot \r(x,t) =  r^2(x, t_*) + r^2(\phi^{t_*}(x), t - t_*) + 2  \r(x,  t_*) \cdot \r(\phi^{t_*}(x), t - t_*) \, ,
\label{expansion}
\eeq
\end{widetext}
that also holds for a particular $x'$. Let us point out that the average~\eqref{constraint} is a conditional average over $x$, and that $x'$ depends on $x$ via the proposal $g(x \rightarrow x')$. We take this into account in the calculation of the left hand side of equation~\eqref{constraint} for the observable $r^2(x; {t})$ expanded as in~\eqref{expansion}:
\begin{widetext}
\begin{eqnarray}
\E[ r^2(x', t)  - r^2(x, t) | x] &=& \E[ r^2(x', t) | x] -  r^2(x, t) = \notag \\ 
&=&  \E[ r^2(x', t_*) |x ] -  r^2(x, t_*)  + \E [r^2(\phi^{t_*}(x'), t - t_*)|x]  -  r^2( \phi^{t_*}(x), t - t_*) \notag \\ 
&+& 2 \E[ \r(x', t_*) \cdot \r( \phi^{t_*}(x'), t - t_*) | x]  -  2 (\r( x, t_*) \cdot \r(\phi^{t_*}(x), t - t_*   )  \, .
\label{fullaverage}
\end{eqnarray}
\end{widetext}
To simplify this expression we make the following assumptions.
Firstly, we assume that up to time $t_*$ the trajectories associated with the flows $\phi^{t_*}(x)$ and $\phi^{t_*}(x')$ are practically indistinguishable and thus on average $  \E[ r^2(t_*, x') |x] = r^2(t_*, x) $. This cancels out the first two terms in equation~\eqref{fullaverage}. Secondly, we assume that on average the dot product of $(\r(t_*, x'), \r(t - t_*, \phi^{t_*}(x')) )$ is equal to that of $(\r(t_*, x), \r(t - t_*, \phi^{t_*}(x))$, thereby simplifying the last two terms in equation~\eqref{fullaverage}. By using the right hand side of the constraint~\eqref{constraint} and combining it with the previous assumptions we get
\beq
\E[r^2( \phi^{t_*}(x'),   t - t_*) | x] -  r^2( \phi^{t_*}(x), t - t_*) = \frac{a - 1}{\beta} \, .
\label{almostfinal}
\eeq

To solve for $t_*$ in terms of $r^2(x, t)$ we need a further simplification. For that, notice that in the diffusive regime we expect that $\E[r^2(x', t - t_*)] = 2 D(t - t_*)$ for an arbitrary $x'$ (this includes  the point $\phi^{t_*}(x')$). And it is reasonable to assume that the following relation
\beq
r^2(\phi^{t_*}(x), t - t_*) \approx \frac{r^2(x, t) (t - t_*)}{t}
\, ,
\eeq
 between the observed value $r^2(x, t) $ and $r^2(\phi^{t_*}(x), t - t_*)$ holds.
By inserting these two approximations into~\eqref{almostfinal} we obtain
\beq
(t - t_*) \left[ 2 D - \frac{r^2_{t}}{t} \right]  = \frac{a - 1}{\beta} \,.
\eeq
Rearranging  terms and solving for $t_*$ we get
\beq
t_* = t - \frac{a - 1}{ 2 D \beta \left( 1 - \frac{r^2}{2D t} \right)}  \, ,
\eeq
with the further simplification of the notation $r^2 := r^2_{t}$. Finally, to guarantee that $t_*$ belongs to the proper interval we impose the following constraint
\beq\label{tsa}
t_* = \max \Bigl\{ t_{\rm min}, t - \bigg \lvert \frac{a-1}{2D \beta (1 -\frac{r^2}{2Dt} ) } \bigg \rvert \Bigr\} \, ,
\eeq
which is the equation~\eqref{maintstar}. The parameter $t_{\rm min}$ is included to guarantee that $t_*$ is physically meaningful (e.g., in most problems $t_* > 0$).
In order to simplify the calculations, the results above were computed using $E(x)=r^2$ as an observable. Nevertheless, the relation between $t_*$ and $r$ in Eq.~\eqref{tsa} is expected to lead also to efficient correlation of the trajectories for the case $E(x)=r$ because there is a monotonic relation between $r$ and $r^2$ (by construction, $r \ge 0$) and because local proposals in $r^2$ will tend to be local in $r$ as well. Our numerical results, obtained for $E(x)=r$,  confirm this expectation.


\end{document}